\documentclass[11pt]{article}
\usepackage[a4paper,margin=1in]{geometry}

\usepackage{amsmath,amssymb}
\usepackage{graphicx}
\usepackage{url}

\usepackage{pgfplots}
\pgfplotsset{compat=1.18}
\usepackage{cite}
\usepackage{tikz}
\usetikzlibrary{positioning,arrows.meta,calc}

\usepackage{algorithm}
\usepackage{algorithmic}

\usepackage{xcolor}

\usepackage[bookmarks=false]{hyperref}

\begin{document}

\title{Semantic Communications in the THz Band}

\author{
Fatima Ismail,
Hadi Sarieddeen, and Jihad Fahs\\
Department of Electrical and Computer Engineering\\
American University of Beirut, Beirut, Lebanon\\
fmi15@mail.aub.edu, hadi.sarieddeen@aub.edu.lb, jihad.fahs@aub.edu.lb%
\thanks{This work is supported by the American University of Beirut's University Research Board and Vertically Integrated Projects Program.}
}

\date{}
\maketitle

\begin{abstract}
Semantic and terahertz (THz)-band communications are algorithmic and spectral enablers of future wireless networks. This work investigates deep learning-based semantic communication (DeepSC) over THz channels. We show that DeepSC models trained solely under additive white Gaussian noise generalize well to the tested THz block- and fast-fading channels when receiver-side compensation is applied. To enable fully data-driven reception, we propose a lightweight neural detector that does not require channel state information (CSI). At \(0.3\)~THz, DeepSC outperforms a throughput-matched traditional coded communication system baseline over \(0\!-\!12\)~dB signal-to-noise (SNR) ratio, achieving more than \(50\) percentage-point higher Bilingual Evaluation Understudy unigram (BLEU-1) score.
 The proposed pilot-free detector outperforms minimum mean square error (MMSE) equalization with both perfect and imperfect CSI and remains robust to frequency offsets up to 50~MHz, highlighting the resilience of semantic communication to THz channel impairments.
\end{abstract}


\section{Introduction}

Future wireless systems must support massive connectivity, ultra-low latency, and extremely high data rates~\cite{rajatheva2020white}. The THz band offers abundant spectrum above \(0.1\)~THz~\cite{9112745,akyildiz2022terahertz}, but suffers from severe path loss, frequency selectivity, and molecular absorption~\cite{9591285,9514889}, making accurate channel state information (CSI) acquisition difficult, especially at low signal-to-noise ratios (SNRs). This challenge is further amplified by THz-specific effects such as mutual coupling, spatial correlation, and near-field propagation~\cite{11006064}.

In parallel, semantic communication transmits meaning rather than raw bits, offering robustness and efficiency at low SNR. Deep learning-based semantic communication (DeepSC)~\cite{xie2021deep} outperforms conventional schemes under additive white Gaussian noise (AWGN) and Rayleigh fading~\cite{xie2021deep}, and has been extended to multimodal~\cite{zhang2022unified}, speech-focused~\cite{weng2021semantic}, and noise-resilient variants~\cite{peng2024robust}. However, these systems are mostly evaluated under simplified channels, leaving their behavior in THz environments largely unexplored, where equalizers such as minimum mean square error (MMSE) detection require accurate CSI that is often unreliable under mobility.

While learning-based detectors such as MMNet~\cite{khani2020adaptive}, OAMP-Net~\cite{he2018model}, and VBINet~\cite{9672704} reduce CSI dependence, they target traditional systems with discrete constellation symbols and typically still rely on pilots or explicit CSI. Recent semantic/joint source-channel coding (JSCC) denoisers mainly target image transmission and operate after channel compensation or under channel-knowledge assumptions~\cite{10436728,10448094,duan2024dm}. Thus, they do not address pilot-free equalization for text semantic communication, where DeepSC transmits continuous channel-symbol sequences with sentence-level dependencies rather than independent constellation symbols. To the best of our knowledge, pilot-free neural equalization for text-based semantic channel symbols has not been studied before. This gap is especially important in THz systems, where pilot overhead and CSI acquisition are costly and unreliable~\cite{9591285}.

Moreover, having the receiver CSI-free is not enough if the full semantic model must be retrained whenever the channel changes. DeepSC and related systems are often trained under specific channel distributions~\cite{xie2021deep,zhang2022unified,weng2021semantic,peng2024robust}, so changing fading conditions require costly retraining or fine-tuning in dynamic fading environments like THz channels.

We address these limitations with a fully data-driven, CSI-free THz semantic communication framework for text data. We propose a lightweight Transformer-based ND that recovers continuous DeepSC channel symbols before decoding, requires no pilots during training or inference, and remains robust to \(\pm50\)~MHz carrier-frequency offsets. Unlike constellation-level detectors~\cite{khani2020adaptive,he2018model,9672704} or image denoisers~\cite{10436728}, the proposed ND targets pilot-free text semantic equalization.

Our main contributions are: (i) we demonstrate that AWGN-trained DeepSC can operate over the tested fading channels without channel-specific retraining when receiver-side compensation is applied; (ii) we propose a pilot-free ND that outperforms MMSE equalization with perfect CSI; and (iii) we analyze MMSE equalization under imperfect CSI, showing that DeepSC is
more robust to CSI errors than traditional coded transmission in the considered
THz channels.

\section{System and Channel Models}
\label{sec:format}

\begin{figure*}[htb!]
    \centering
    \includegraphics[width=1\linewidth]{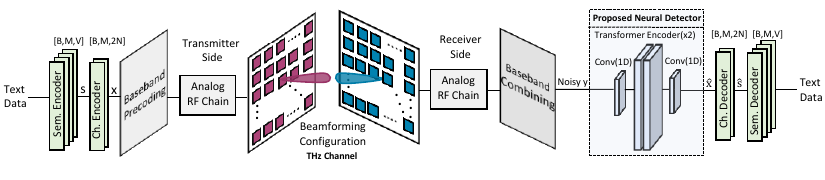}
    \caption{The DeepSC framework augmented with the proposed ND in a THz MIMO system.}
    \label{fig:my_figure}
\end{figure*}

We consider an end-to-end semantic communication system based on DeepSC~\cite{xie2021deep}, which transmits the semantic meaning of a sentence \(\mathbf{s}=[w_1,\dots,w_L]\). The transmitter includes a semantic encoder \(S_{\beta}(\cdot)\) and a channel encoder \(C_{\alpha}(\cdot)\), parameterized by \(\beta\) and \(\alpha\), respectively. The semantic encoder extracts semantic features, and the channel encoder maps these features to a symbol vector \(\mathbf{x}=C_{\alpha}(S_{\beta}(\mathbf{s})) \in \mathbb{C}^{M\times1}\) for transmission over a THz channel with coherence interval \(M\). After \(M\) transmissions, the received signal vector is
\(\mathbf{y}=h\mathbf{x}+\mathbf{n}\), where \(h\) denotes the
effective scalar channel obtained by projecting the THz multiple-input multiple-output (MIMO) channel onto
the selected transmit and receive analog beamforming vectors, and
\(\mathbf{n}\sim\mathcal{CN}(\mathbf{0},\sigma_n^2\mathbf{I}_M)\) is AWGN. The instantaneous per-symbol SNR for a given channel realization $h$ is defined as $\mathrm{SNR}=\frac{P|h|^2}{\sigma_n^2}$, where $P=\mathbb{E}[|x_m|^2]$ denotes the average transmit power per symbol.
 At the receiver, the channel decoder \(C^{-1}_{\delta}(\cdot)\) and the semantic decoder \(S^{-1}_{\chi}(\cdot)\) reconstruct the transmitted sentence as \(\hat{\mathbf{s}}=S^{-1}_{\chi}(C^{-1}_{\delta}(\mathbf{y}))\).

We adopt realistic THz channel realizations generated using the TeraMIMO simulator~\cite{9591285}, which follows a Saleh--Valenzuela--type clustered multipath model and incorporates frequency-dependent molecular absorption based on high-resolution transmission molecular absorption database (HITRAN) data, together with subarray (SA) geometry. In this architecture, the transmitter and receiver consist of an SA composed of multiple antenna elements. 
At subcarrier index \(k\), the scalar beamformed coefficient between the \(q^{(t)}\)-th transmit SA and the \(q^{(r)}\)-th receive SA is
\begin{equation}
\begin{aligned}
{h}_{q^{(r)},q^{(t)}}[k]
&=
{\mathbf a}^{(r)T}\!\Big(
\boldsymbol{\Phi}_{0}^{(r)}{}_{q^{(r)},\,q^{(r)}}
\Big)\,
\mathbf{H}_{q^{(r)},q^{(t)}}[k] \\
&\quad\times
{\mathbf a}^{(t)}\!\Big(
\boldsymbol{\Phi}_{0}^{(t)}{}_{q^{(t)},\,q^{(t)}}
\Big),
\end{aligned}
\end{equation}
where \({\mathbf a}^{(t)}(\cdot)\) and \({\mathbf a}^{(r)}(\cdot)\) denote the transmit and receive beamforming vectors associated with the selected steering directions. The matrix \(\mathbf{H}_{q^{(r)},q^{(t)}}[k]\) represents the frequency-domain THz MIMO channel, incorporating LoS and NLoS components, antenna gains, array responses, propagation delays, and molecular absorption effects, as modeled in~\cite{9591285}.

\section{Proposed Detector and Training Methods}
\label{sec:pagestyle}

\subsection{Neural detector architecture}

To avoid explicit THz channel estimation, we replace conventional equalization with a lightweight pilot-free ND before the channel decoder (Fig.~\ref{fig:my_figure}). Following DeepSC~\cite{xie2021deep}, the received sequence is reshaped as \([B,M,2]\), where the last dimension stores the real and imaginary parts, with \(M=NL\), where \(N\) is the number of complex symbols per word, and \(L=30\) is the sentence length.

The ND uses a 1D convolution (\(k=3\), padding \(1\)) to project the input to \(d_{\mathrm{model}}=64\), capturing local symbol-level distortions at low complexity. Fixed sinusoidal positional encoding preserves symbol order, after which two Transformer encoder layers with \(8\) attention heads and feed-forward network (FFN) dimension \(256\) exploit global relations between words in a sentence; since channel-coded symbols inherit dependencies from the input sentence, self-attention allows each position to exploit the full received sequence rather than denoising each symbol independently. A final 1D convolution (\(k=3\), padding \(1\)) maps the output back to two dimensions, yielding a denoised signal with THz channel effects implicitly removed. The detector has \(100{,}802\) trainable parameters, which is compact for a Transformer-based sequence receiver~\cite{saleem2024transrx,sun2020mobilebert}, and avoids CSI estimation, pilot processing, and channel-dependent equalizer computation, supporting its lightweight design.

\subsection{Overview of the Training Strategy}

\subsubsection{AWGN training rationale}
\label{sec:awgn_siso_rationale}

We motivate AWGN training from the post-equalization view. Under perfect-CSI equalization, the fading channel \(y=hx+n\) becomes $\hat{x}_{\mathrm{MMSE}}=\alpha x+\tilde{n}$, where \(\alpha\) is a channel-dependent gain and \(\tilde{n}\) remains white Gaussian. Thus, the decoder mainly sees an additive Gaussian perturbation, as in AWGN training, with scaled noise variance. This motivates training DeepSC over \(y=x+n\), \(n\sim\mathcal{CN}(0,\sigma_n^2)\). Since MMSE requires perfect CSI in this case, we replace it with our proposed ND, enabling a fully data-driven system that recovers the clean channel-symbol sequence from THz-corrupted observations without pilots or explicit CSI.

\subsubsection{Overview of the training strategy}

We adopt a two-stage training procedure.

\textit{Stage 1: DeepSC pretraining over AWGN.}
The DeepSC encoder–decoder is trained end-to-end over an AWGN channel to optimize semantic transmission by minimizing
\begin{equation}
\mathcal{L}_{\text{total}}
= \mathcal{L}_{\text{CE}}(\mathbf{s},\hat{\mathbf{s}};\alpha,\beta,\chi,\delta)
- \lambda\,\mathcal{L}_{\text{MI}}(\mathbf{x},\mathbf{y};T,\alpha,\beta).
\end{equation}

where $\mathcal{L}_{\text{CE}}$ measures semantic distortion and $\mathcal{L}_{\text{MI}}$ encourages a higher achievable rate. The weighting factor is set to $\lambda=0.0009$ following the DeepSC framework~\cite{xie2021deep}. Training is performed for approximately 75 epochs at 5–10\,dB SNR, after which $S_{\beta}$, $C_{\alpha}$, $C^{-1}_{\delta}$, and $S^{-1}_{\chi}$ are frozen.

\textit{Stage 2: ND training over THz channels.}
With DeepSC frozen, an ND \(D_{\boldsymbol{\theta}}(\cdot)\) is inserted before the channel decoder and trained to map THz-corrupted observations to the clean transmitted channel-symbol sequence, \(\hat{\mathbf{x}}=D_{\boldsymbol{\theta}}(\mathbf{y})\). 

It minimizes \(\mathcal{L}_{\mathrm{MSE}}=\|\hat{\mathbf{x}}-\mathbf{x}\|^2\), using \(\mathbf{x}\) as the supervised target. This directly aligns the ND output with the clean channel-symbol distribution expected by the fixed AWGN-trained DeepSC decoder, enabling CSI-free recovery without pilots, explicit channel estimation, or matrix inversions.

Training uses Adam~\cite{article} for \(80\) epochs with initial learning rate \(10^{-4}\) and cosine annealing~\cite{10104453}, as summarized in Algorithm~\ref{alg:training}.

\begin{algorithm}[t]
\caption{Training of DeepSC with ND}
\label{alg:training}
\begin{algorithmic}[1]
\STATE \textbf{Stage 1: AWGN pretraining.} Train DeepSC end-to-end over an AWGN channel using \(\mathcal{L}_{\text{total}}\), then freeze \((\alpha,\beta,\delta,\chi)\).
\STATE \textbf{Stage 2: THz ND training.}
\FOR{epoch \(=1\) to \(N\)}
    \STATE Encode \(\mathbf{s}\) as \(\mathbf{x}=C_{\alpha}(S_{\beta}(\mathbf{s}))\), and generate \(\mathbf{y}=h\mathbf{x}+\mathbf{n}\) over the THz channel.
    \STATE Estimate \(\hat{\mathbf{x}}=D_{\boldsymbol{\theta}}(\mathbf{y})\) and update \(\boldsymbol{\theta}\) by minimizing \(\|\hat{\mathbf{x}}-\mathbf{x}\|^2\).
\ENDFOR
\end{algorithmic}
\end{algorithm}

\section{Main Results and Discussion}
We first use perfect-CSI MMSE as a controlled reference, 
then evaluate the fully CSI-free system where the proposed ND replaces MMSE 
while keeping the DeepSC encoder-decoder fixed after AWGN training. Symbol-level NDs 
are omitted as they assume discrete constellations, whereas our transmitter 
uses continuous embeddings.

\emph{Simulation settings:}
We consider an indoor THz scenario at \(0.3\)~THz over \(10\)~GHz bandwidth, consistent with TeraMIMO simulation settings and prior indoor THz studies~\cite{9591285,10123397,9839013}. The path-loss exponent is \(2\)~\cite{9473756}, and the noise variance is normalized to one. Owing to the short wavelength, we use half-wavelength antenna spacing~\cite{8311993} and \(4096\times4096\) antenna elements at the transmitter and receiver~\cite{faisal2020ultramassive}, enabled by compact THz array technologies~\cite{AKYILDIZ201646}. The THz channel is generated using TeraMIMO with its default indoor Saleh--Valenzuela parameters~\cite{9591285}: LoS plus multipath propagation, HITRAN absorption, \(T=298.15\)~K, \(p=1\)~atm, ideal geometry-aligned analog beamforming, exponential cluster/ray arrivals with rates \(0.13\) and \(0.37~\mathrm{ns}^{-1}\), and Poisson-distributed cluster/ray counts. Simulations use Europarl (80k/10k train/test sentences). The ND is trained 
under block fading with a new independent THz channel and noise realization 
per batch of 64 sentences. Reported curves average \(30\) independent full-test passes with newly generated per-batch realizations; across all SNR points, the maximum sample standard deviation is \(1.3\times10^{-2}\) BLEU-1 and the \(95\%\) confidence-interval half-widths remain below \(5\times10^{-3}\) BLEU-1.

\begin{figure}[!t]
\centering

\begin{minipage}[t]{0.48\textwidth}
\centering
\begin{tikzpicture}
    \begin{axis}[
      width=\linewidth,
      height=6cm,
      xlabel={SNR (dB)},
      ylabel={\footnotesize{BLEU score (1-gram)}},
      grid=major,
      tick label style={font=\footnotesize},
      line width=0.5pt,
      mark size=1.5pt,
      xmin=0, xmax=21,
      ymin=0, ymax=1,
      xtick={0,3,...,21},
      ytick={0,0.2,...,1},
      legend style={
        draw=none,
        fill=white,
        fill opacity=0.7,
        text opacity=1,
        font=\scriptsize,
        cells={anchor=west},
        at={(1,0.77)},
        anchor=north east,
        legend columns=1,
        row sep=-0.5pt,
        nodes={inner ysep=1pt}
      }]
      
      \addplot[black, dashed, mark=*, mark options={solid, fill=white, draw=black},line width=0.8pt] 
        table[x=SNR, y=AWGN_AWGN, col sep=comma] {channel_comparison.csv};
      \addlegendentry{AWGN (AWGN)}

      \addplot[blue, dashed, mark=square*, mark options={solid, fill=white, draw=blue},line width=0.8pt] 
        table[x=SNR, y=Rayleigh_AWGN, col sep=comma] {channel_comparison.csv};
      \addlegendentry{Rayleigh (AWGN)}

      \addplot[red, dashed, mark=triangle*, mark options={solid, fill=white, draw=red},line width=0.8pt] 
        table[x=SNR, y=Rician_AWGN, col sep=comma] {channel_comparison.csv};
      \addlegendentry{Rician (AWGN)}

      \addplot[green!70!black, dashed, mark=diamond*, mark options={solid, fill=white, draw=green!70!black},line width=0.8pt] 
        table[x=SNR, y=THz_AWGN, col sep=comma] {channel_comparison.csv};
      \addlegendentry{THz (AWGN)}

      \addplot[orange!60!red!80!brown, dashed, mark=x, mark options={solid, draw=orange!60!red!80!brown},line width=0.8pt] 
        table[x=SNR, y=AWGN_THzMol, col sep=comma] {channel_comparison.csv};
      \addlegendentry{THz+Mol (AWGN)}
      
      \addplot[blue, solid, mark=square*, mark options={solid, fill=blue},line width=0.8pt] 
        coordinates {
          (0, 0.52896722)
          (3, 0.71212312)
          (6, 0.78166691)
          (9, 0.85989758)
          (12, 0.87701287)
          (15, 0.90600748)
          (18, 0.91562927)
          (21, 0.92610218)
        };
      \addlegendentry{Rayleigh (Rayleigh)}
      
      \addplot[red, solid, mark=triangle*, mark options={solid, fill=red},line width=0.8pt] 
        coordinates {
          (0, 0.545)
          (3, 0.76)
          (6, 0.86)
          (9, 0.89)
          (12, 0.92)
          (15, 0.925)
          (18, 0.925)
          (21, 0.928)
        };
      \addlegendentry{Rician (Rician)}
      
      \addplot[green!70!black, solid, mark=diamond*, mark options={solid, fill=green!70!black},line width=0.8pt] 
        table[x=SNR, y=THz_THz, col sep=comma] {channel_comparison.csv};
      \addlegendentry{THz (THz)}

      \addplot[orange!60!red!80!brown, solid, mark=x, mark options={solid, fill=orange!60!red!80!brown},line width=0.8pt] 
        table[x=SNR, y=THz_THzMol, col sep=comma] {channel_comparison.csv};
      \addlegendentry{THz+Mol (THz)}
    \end{axis}
\end{tikzpicture}

{\normalsize (a)}
\end{minipage}
\hfill
\begin{minipage}[t]{0.48\textwidth}
\centering
\begin{tikzpicture}
    \begin{axis}[
      width=\linewidth,
      height=6cm,
      xlabel={SNR (dB)},
      ylabel={\footnotesize{SBERT score}},
      grid=major,
      tick label style={font=\footnotesize},
      line width=0.5pt,
      mark size=1.5pt,
      xmin=0, xmax=21,
      ymin=0, ymax=1,
      xtick={0,3,...,21},
      ytick={0,0.2,...,1},
      legend style={
        draw=none,
        fill=white,
        fill opacity=0.7,
        text opacity=1,
        font=\scriptsize,
        cells={anchor=west},
        at={(1,0.77)},
        anchor=north east,
        legend columns=1,
        row sep=-0.5pt,
        nodes={inner ysep=1pt}        
      }]
      
      \addplot[black, dashed, mark=*, mark options={solid, fill=white, draw=black},line width=0.8pt] 
        table[x=SNR, y=AWGN_AWGN, col sep=comma] {bert.csv};
      \addlegendentry{AWGN (AWGN)}

      \addplot[blue, dashed, mark=square*, mark options={solid, fill=white, draw=blue},line width=0.8pt] 
        table[x=SNR, y=Rayleigh_AWGN, col sep=comma] {bert.csv};
      \addlegendentry{Rayleigh (AWGN)}

      \addplot[red, dashed, mark=triangle*, mark options={solid, fill=white, draw=red},line width=0.8pt] 
        table[x=SNR, y=Rician_AWGN, col sep=comma] {bert.csv};
      \addlegendentry{Rician (AWGN)}

      \addplot[green!70!black, dashed, mark=diamond*, mark options={solid, fill=white, draw=green!70!black},line width=0.8pt] 
        table[x=SNR, y=THz_AWGN, col sep=comma] {bert.csv};
      \addlegendentry{THz (AWGN)}

      \addplot[orange!60!red!80!brown, dashed, mark=x, mark options={solid, draw=orange!60!red!80!brown},line width=0.8pt] 
        table[x=SNR, y=AWGN_THzMol, col sep=comma] {bert.csv};
      \addlegendentry{THz+Mol (AWGN)}
      
      \addplot[blue, solid, mark=square*, mark options={solid, fill=blue},line width=0.8pt] 
        coordinates {
          (0, 0.58)
          (3, 0.77)
          (6, 0.82)
          (9, 0.84)
          (12, 0.89)
          (15, 0.90)
          (18, 0.90)
          (21, 0.90)
        };
      \addlegendentry{Rayleigh (Rayleigh)}
      
      \addplot[red, solid, mark=triangle*, mark options={solid, fill=red},line width=0.8pt] 
        coordinates {
          (0, 0.6)
          (3, 0.8)
          (6, 0.84)
          (9, 0.86)
          (12, 0.90)
          (15, 0.91)
          (18, 0.91)
          (21, 0.91)
        };
      \addlegendentry{Rician (Rician)}
      
      \addplot[green!70!black, solid, mark=diamond*, mark options={solid, fill=green!70!black},line width=0.8pt] 
        table[x=SNR, y=THz_THz, col sep=comma] {bert.csv};
      \addlegendentry{THz (THz)}

      \addplot[orange!60!red!80!brown, solid, mark=x, mark options={solid, fill=orange!60!red!80!brown},line width=0.8pt] 
        table[x=SNR, y=THz_THzMol, col sep=comma] {bert.csv};
      \addlegendentry{THz+Mol (THz)}
    \end{axis}
\end{tikzpicture}

{\normalsize (b)}
\end{minipage}

\vspace{0.55cm}

\begin{minipage}[!htbp]{0.5\textwidth}
\centering
\begin{tikzpicture}
    \begin{axis}[
      width=\linewidth,
      height=6cm,
      line width=0.5pt,
      xlabel={SNR (dB)},
      ylabel={\footnotesize{BLEU score (1-gram)}},
      grid=major,
      tick label style={font=\footnotesize},
      xmin=0, xmax=21,
      ymin=0, ymax=1,
      xtick={0,3,...,21},
      ytick={0,0.2,...,1},
      legend style={
        at={(0.99,0.1)},
        anchor=south east,
        font=\scriptsize,
        draw=none,
        fill=white,
        fill opacity=0.75,
        text opacity=1,
        row sep=0.3pt
      }]
      
      \addplot[black, mark=*, line width=0.8pt] 
        table[x=SNR, y=AWGN, col sep=comma] {bleu_data.csv};
        
      \addplot[blue, mark=square*, line width=0.8pt] 
        table[x=SNR, y=T_Distribution, col sep=comma] {bleu_data.csv};
        
      \addplot[green!70!black, mark=diamond*, line width=0.8pt] 
        table[x=SNR, y=Laplacian, col sep=comma] {bleu_data.csv};

      \addplot[orange!60!red!80!brown, mark=x, line width=0.8pt, solid] coordinates {
        (0,0.536) (3,0.835) (6,0.922) (9,0.935)
        (12,0.945) (15,0.945) (18,0.945) (21,0.945)
      };

      \addplot[orange!60!red!80!brown, mark=x, mark options={solid},line width=0.8pt, dotted] coordinates {
        (0,0.53192268) (3,0.83088317) (6,0.92178501) (9,0.93985017)
        (12,0.94198261) (15,0.942352) (18,0.94236033) (21,0.94244818)
      };

      \addplot[orange!60!red!80!brown, mark=x, line width=0.8pt, dashed] coordinates {
        (0,0.53440612) (3,0.82099746) (6,0.92096109) (9,0.93986182)
        (12,0.94213821) (15,0.94235303) (18,0.94252599) (21,0.94255228)
      };

      \legend{
        AWGN,
        T-Distribution,
        Laplacian,
        THz-per batch,
        THz-per sentence,
        THz-per word
      }
    \end{axis}
\end{tikzpicture}

{\normalsize (c)}
\end{minipage}

\caption{(a) BLEU and (b) SBERT scores comparing AWGN-trained models (dashed lines) against same-channel trained models (solid lines) over various fading channels; (c) BLEU scores for AWGN-trained DeepSC under different additive noise distributions and for THz channel under different update rates.}
\label{fig:combined_results}
\end{figure}
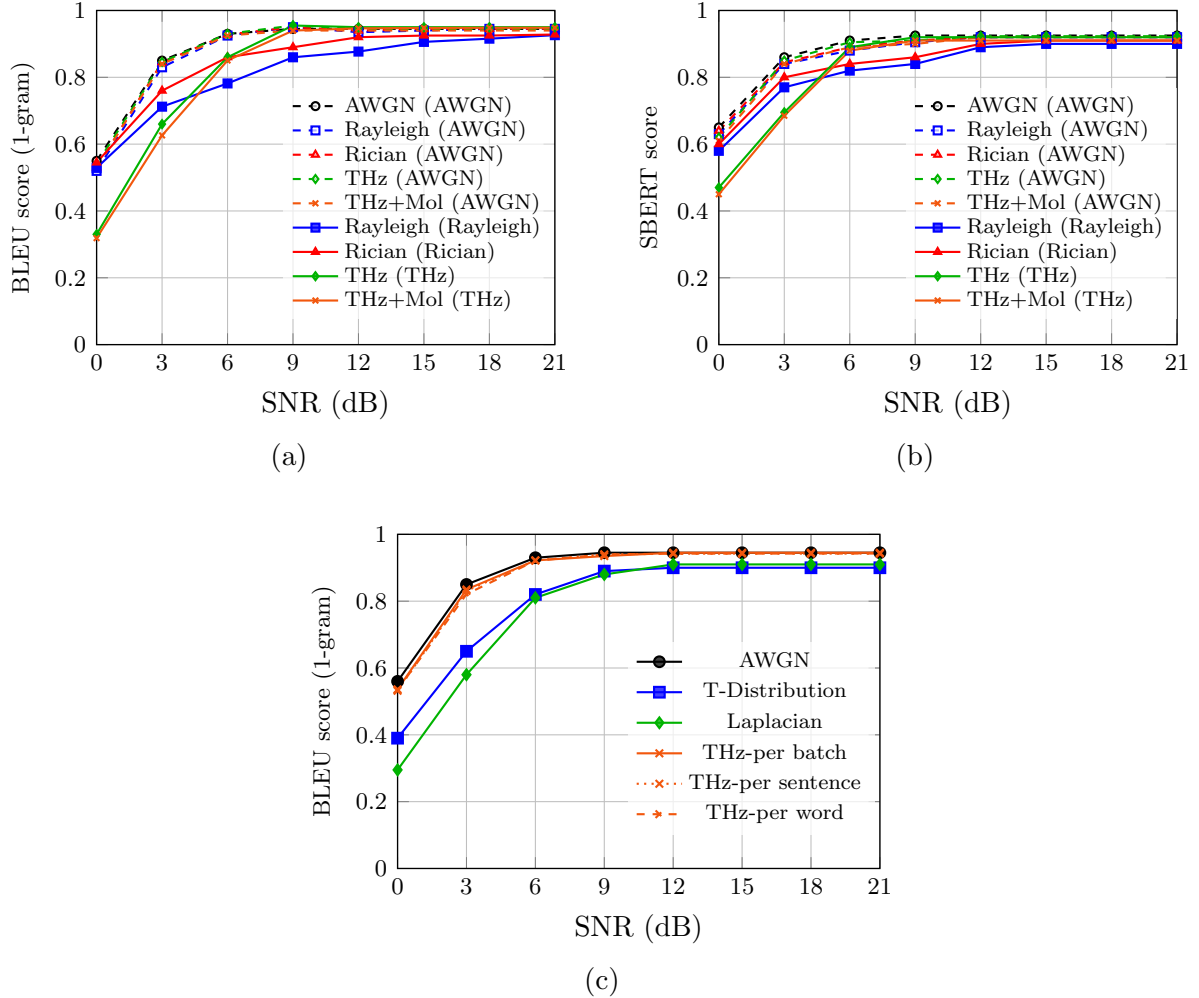

\begin{figure}[t]
\centering

\begin{minipage}[t]{0.48\textwidth}
\centering
\begin{tikzpicture}
    \begin{axis}[
      width=\linewidth,
      height=6cm,
      xlabel={SNR (dB)},
      ylabel={\footnotesize BLEU Score (1-gram)},
      grid=major,
      tick label style={font=\footnotesize},
      line width=0.5pt,
      mark size=1.5pt,
      xmin=0, xmax=21,
      ymin=0, ymax=1,
      xtick={0,3,...,21},
      ytick={0,0.2,...,1},
      legend style={
        at={(0.97,0.5)},
        anchor=south east,
        font=\scriptsize,
        draw=none,
        fill=white,
        fill opacity=0.7,
        text opacity=1,
        row sep=-1.5pt,
        nodes={inner ysep=1pt}
      }]

      \addplot[blue, mark=triangle*, line width=0.8pt] coordinates {
        (0,0.569) (3,0.847) (6,0.92) (9,0.937)
        (12,0.94) (15,0.9385) (18,0.941) (21,0.938)
      };

      \addplot[blue, dashed, mark=triangle*, line width=0.8pt] coordinates {
        (0,0.2557) (3,0.3026) (6,0.3361) (9,0.3810)
        (12,0.4128) (15,0.4807) (18,0.4971) (21,0.5417)
      };

      \addplot[blue, mark=triangle*, dotted, line width=0.8pt] coordinates {
        (0,0.0000) (3,0.0287) (6,0.0410) (9,0.0714)
        (12,0.0905) (15,0.1222) (18,0.1603) (21,0.1858)
      };

      \addplot[black, mark=*, line width=0.8pt] coordinates {
        (0,0.53) (3,0.82) (6,0.90) (9,0.92)
        (12,0.935) (15,0.935) (18,0.935) (21,0.935)
      };

      \addplot[brown, mark=square*, line width=0.8pt] coordinates {
        (0,0) (3,0) (6,0) (9,0)
        (12,0.15) (15,0.995) (18,1) (21,1)
      };

      \legend{Proposed ND, MLP-128, Conv.only, MMSE, Traditional+MMSE}
    \end{axis}
\end{tikzpicture}

{\normalsize (a)}
\end{minipage}
\hfill
\begin{minipage}[t]{0.48\textwidth}
\centering
\begin{tikzpicture}
  \begin{axis}[
    width=\linewidth,
    height=6cm,
    xlabel={SNR (dB)},
    ylabel={\footnotesize BLEU Score (1-gram)},
    grid=major,
    tick label style={font=\footnotesize},
    line width=0.5pt,
    mark size=1.5pt,
    xmin=0, xmax=21,
    ymin=0, ymax=1,
    xtick={0,3,...,21},
    ytick={0,0.2,...,1},
    legend style={
      at={(0.98,0.25)},
      anchor=south east,
      font=\scriptsize,
      draw=none,
      fill=white,
      fill opacity=0.7,
      text opacity=1,
      legend cell align=left,
      legend columns=1,
      row sep=-0.5pt,
      nodes={inner ysep=1pt}
    }
  ]

    \addplot[blue, solid, mark=triangle*, line width=0.8pt]
      table[x=SNR, y=Neural_Detector, col sep=comma] {bleu_data_new.csv};
    \addlegendentry{Proposed ND}
    
    \addplot[black, solid, mark=*, line width=0.8pt]
      table[x=SNR, y=MMSE_Full_CSI, col sep=comma] {bleu_data_new.csv};
    \addlegendentry{MMSE Full CSI}

    \addplot[green!70!black, solid, mark=x, line width=0.8pt]
      table[x=SNR, y=MMSE_Est_CSI, col sep=comma] {bleu_data_new.csv};
    \addlegendentry{MMSE Est.\ CSI}

    \addplot[violet, dashed, mark=triangle*, line width=0.8pt] coordinates {
      (0, 0.2815) (3, 0.4860) (6, 0.6348) (9, 0.7027)
      (12, 0.7518) (15, 0.7539) (18, 0.7743) (21, 0.7719)
    };
    \addlegendentry{ND offset}

    \addplot[gray, dashed, mark=*, line width=0.8pt] coordinates {
      (0, 0.13) (3, 0.13) (6, 0.125) (9, 0.116)
      (12, 0.113) (15, 0.118) (18, 0.104) (21, 0.105)
    };
    \addlegendentry{MMSE offset}

    \draw[black, line width=0.8pt] (axis cs:1.7,0.25) ellipse [x radius=1.2, y radius=0.18];
    \draw[black, thick, ->, >=Stealth]
      (axis cs:2.7,0.36) -- (axis cs:4.0,0.25)
      node[right,yshift=-4pt, font=\tiny, align=left]{$\pm50\,\mathrm{MHz}$ offset};

  \end{axis}
\end{tikzpicture}

{\normalsize (b)}
\end{minipage}

\vspace{0.55cm}

\begin{minipage}[t]{0.50\textwidth}
\centering
\begin{tikzpicture}
  \begin{axis}[
    width=\linewidth,
    height=6cm,
    xlabel={SNR (dB)},
    ylabel={\footnotesize BLEU Score (1-gram)},
    grid=major,
    tick label style={font=\footnotesize},
    line width=0.5pt,
    mark size=1.5pt,
    xmin=0, xmax=21,
    ymin=0, ymax=1,
    xtick={0,3,...,21},
    ytick={0,0.2,...,1},
    legend style={
      at={(0.97,0.18)},
      anchor=south east,
      font=\scriptsize,
      draw=none,
      fill=white,
      fill opacity=0.7,
      text opacity=1,
      cells={anchor=west},
      legend columns=1,
      row sep=-0.5pt,
      nodes={inner ysep=1pt}
    }
  ]

    \addplot[blue, mark=triangle*, line width=0.8pt]
      table[x=SNR, y=DeepSC_Neural_01, col sep=comma] {bleu_alpha_data.csv};
    \addlegendentry{Proposed ND}

    \addplot[blue, mark=triangle*, line width=0.8pt, dashed, forget plot]
      table[x=SNR, y=DeepSC_Neural_05, col sep=comma] {bleu_alpha_data.csv};

    \addplot[black, mark=*, line width=0.8pt]
      table[x=SNR, y=DeepSC_MMSE_01, col sep=comma] {bleu_alpha_data.csv};
    \addlegendentry{MMSE}

    \addplot[black, mark=*, line width=0.8pt, dashed, forget plot]
      table[x=SNR, y=DeepSC_MMSE_05, col sep=comma] {bleu_alpha_data.csv};

    \addplot[brown, mark=square*, line width=0.8pt] coordinates {
      (0,0) (3,0) (6,0) (9,0) (12,0) (15,0) (18,0.08) (21,0.12)
    };
    \addlegendentry{Traditional+MMSE}

    \addplot[brown, mark=square*, line width=0.8pt, dashed, forget plot] coordinates {
      (0,0) (3,0) (6,0) (9,0) (12,0) (15,0) (18,0) (21,0)
    };

  \end{axis}
\end{tikzpicture}

{\normalsize (c)}
\end{minipage}

\caption{BLEU-1 over THz channels: (a) DeepSC with proposed/ablated NDs and MMSE, compared with throughput-matched LDPC-coded 64-QAM+MMSE; (b) proposed ND versus MMSE with full/pilot CSI and \(\pm50\,\mathrm{MHz}\) offset; (c) noisy-CSI robustness, with solid/dashed lines for \(\alpha=0.1/0.5\).}
\label{fig:three_plots}
\end{figure}
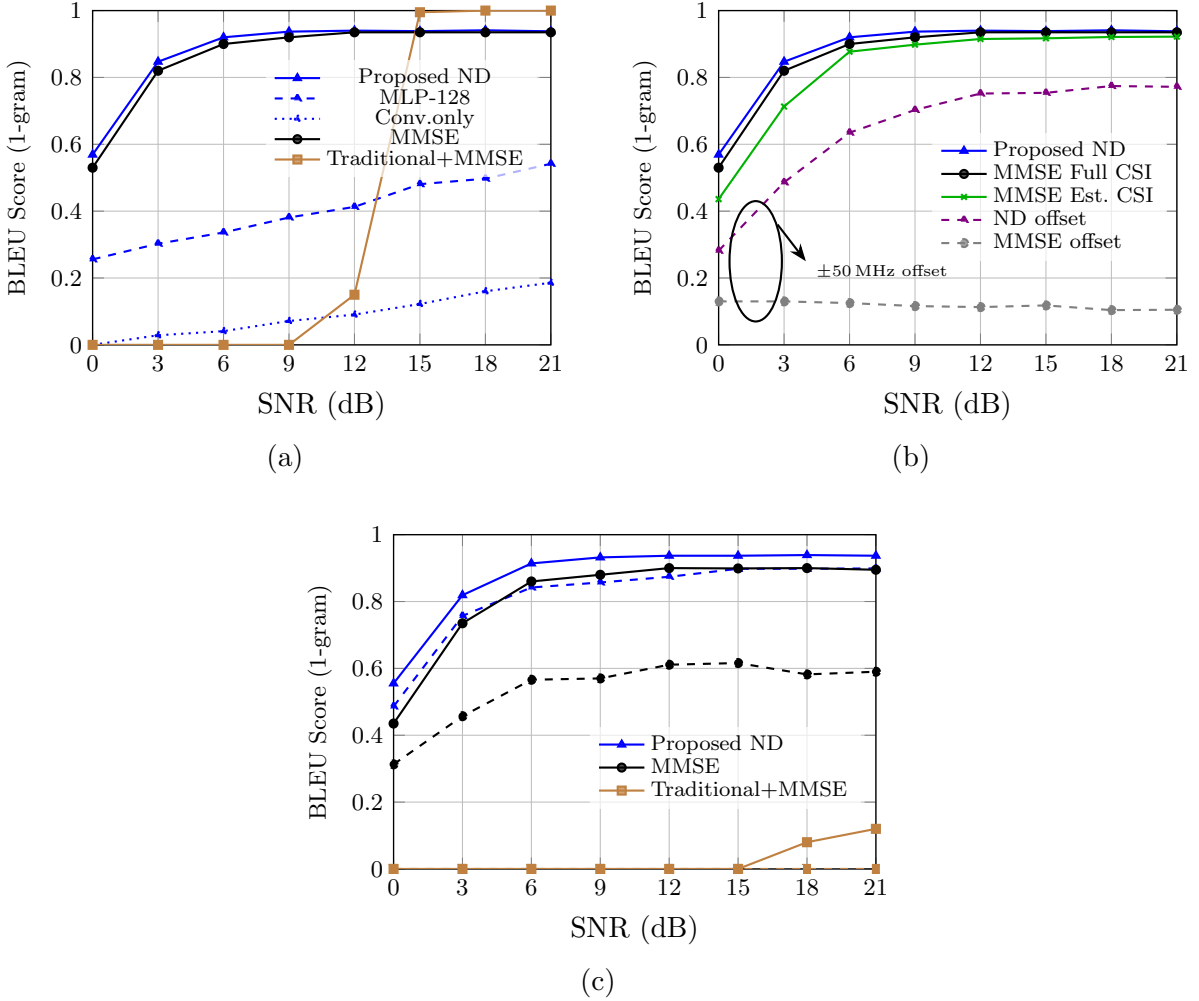

\subsection{Generalization Across Channel Types}
We first consider block fading, where the channel changes per batch of \(64\) 
sentences, using MMSE with perfect CSI as a controlled reference to validate 
the post-equalization noise argument of Section~\ref{sec:awgn_siso_rationale} 
before replacing it with the pilot-free ND in Sections~\ref{baselines} and~\ref{performance}.
Performance is evaluated using BLEU-1~\cite{papineni2002bleu} and 
SBERT~\cite{reimers2019sentence}. Fig.~\ref{fig:combined_results}(a,b) compare AWGN-trained 
DeepSC tested over different fading channels with models trained and tested on 
the same channel. The AWGN-trained model matches or exceeds Rayleigh/Rician-trained 
models on their respective channels and outperforms THz-trained models under all 
THz settings, especially at low SNR (\(0\!-\!6\)~dB). Thus, AWGN-only training 
provides strong cross-channel generalization without channel-specific retraining. 
We further test frequency-selective THz molecular absorption at 
\(f_c = 445\)~GHz, where water vapor causes a strong absorption 
peak~\cite{5995306}. Despite severe attenuation, the AWGN-trained model 
maintains high BLEU-1 and SBERT, confirming that generalization is governed 
by the post-equalization noise statistics rather than the specific fading profile.

To further evaluate the proposed training strategy under fast-fading conditions, we 
simulate THz channels that vary per batch, sentence, and word. For a test set 
containing \(10\)k sentences, word-level fading corresponds to approximately 
\(3\times10^5\) independent THz channel realizations. As shown in 
Fig.~\ref{fig:combined_results}(c), the AWGN-trained model maintains nearly 
identical BLEU-1 performance across the three fading update rates, showing that 
the fading variation rate has limited impact after equalization. This is 
consistent with Section~\ref{sec:awgn_siso_rationale}: under perfect CSI, MMSE 
maps the SISO fading channel to an equivalent additive-noise channel, so the 
decoder observes the post-equalization perturbation rather than the fading law 
itself.

To verify that the additive perturbation is the dominant factor, we test AWGN-trained 
DeepSC under different noise types: zero-mean Gaussian noise with variance $\sigma_n^2$, zero-mean Laplace noise with scale $b=\sqrt{\sigma_n^2/2}$, and Student-$t$ noise with $\nu=3$ degrees of freedom and scale chosen to match variance $\sigma_n^2$. Fig.~\ref{fig:combined_results}(c) 
also shows that, unlike the fading update rate, changing the noise distribution 
noticeably degrades BLEU-1; SBERT follows the same trend. This confirms that 
semantic performance is more sensitive to the post-equalization noise distribution 
than to the fading profile, motivating the pilot-free ND evaluated next as the practical CSI-free alternative.

\subsection{Baseline Comparison and Receiver Ablations}
\label{baselines}
To ensure a fair comparison with a traditional coded communication system, we implement a Sionna-based pipeline~\cite{hoydis2022sionna} with ASCII source encoding, LDPC coding, \(64\)-QAM modulation, and MMSE equalization. Since DeepSC uses \(8\) complex channel symbols per word and Europarl has an average of \(4.89\) characters/word, ASCII encoding gives an effective source rate of \((4.89\times 8)/8=4.89\) bits/channel symbol. We therefore use LDPC-coded \(64\)-QAM with \(R=4/5\), yielding \(6R=4.8\) information bits/channel symbol. As shown in Fig.3(a), DeepSC with perfect-CSI MMSE outperforms this throughput-matched baseline over \(0\!-\!12\)~dB under the same THz channel. The proposed pilot-free ND matches perfect-CSI MMSE with DeepSC and outperforms two simpler denoisers: a Convolution-only variant that removes the Transformer layers, and a two-hidden-layer token-wise MLP with 128 neurons/layer, confirming the benefit of Transformer self-attention for sequence-aware equalization in text-based semantic communication.

\subsection{Performance of the Proposed ND}
\label{performance}

We evaluate the proposed ND on noisy \(300\)~GHz THz transmissions and compare it with MMSE using full CSI and pilot-based CSI estimation with \(10\%\) pilots per coherence block, estimated by least squares~\cite{ls}. As shown in Fig.~\ref{fig:three_plots}(b), the ND outperforms both MMSE variants, especially at low SNR, despite requiring no CSI during training or inference. To test frequency generalization, the ND trained at \(300\)~GHz is evaluated without retraining at \(300.05\)~GHz, corresponding to a \(50\)~MHz offset. It remains robust and outperforms MMSE with mismatched CSI, where the \(300\)~GHz CSI is applied to \(300.05\)~GHz data, although full-CSI MMSE at the offset frequency remains the upper reference.

We further study imperfect CSI for MMSE equalization in both DeepSC and the throughput-matched LDPC-coded \(64\)-QAM baseline, and compare them with the proposed ND. The noisy estimate is modeled as \(\hat{h}=h+\Delta h\), where \(\Delta h\sim\mathcal{CN}(0,\alpha |h|^2)\), with \(\alpha=0.1\) and \(\alpha=0.5\). As shown in Fig.~\ref{fig:three_plots}(c), the proposed ND remains robust under CSI-related distortions although it is trained only on \(300\)~GHz THz observations. The effect of CSI mismatch under MMSE can be understood from
\begin{equation}
\hat{\mathbf{x}}
=
\tfrac{\hat{h}^*h}{|\hat{h}|^2+\sigma_n^2}\mathbf{x}
+
\tfrac{\hat{h}^*}{|\hat{h}|^2+\sigma_n^2}\mathbf{n}
\stackrel{\text{high SNR}}{\approx}
\tfrac{h}{\hat{h}}\mathbf{x}
+
\tfrac{\mathbf{n}}{\hat{h}}
\approx
\bigl(1-\tfrac{\Delta h}{h}\bigr)
\bigl(\mathbf{x}+\tfrac{\mathbf{n}}{h}\bigr),
\end{equation}
where \(\Delta h\ll h\). Thus, CSI error introduces a complex multiplicative distortion on the equalized signal \(\mathbf{x}+\mathbf{n}/h\). In LDPC-coded \(64\)-QAM, such gain and phase errors directly perturb constellation decisions and degrade text recovery. In contrast, DeepSC's continuous high-dimensional representations make the decoded meaning less sensitive to moderate perturbations. The ND further generalizes beyond training, remaining robust under noisy CSI by bypassing corrupted channel estimates and recovering channel-symbol sequences directly.

\emph{Latency Analysis:} On a single NVIDIA V100 GPU allocation (32~GB, v100d32q) with inference batch size \(64\), processing the full \(10\)k-sentence test set requires \(308\)~s with perfect-CSI MMSE, and around \(314.7\)~s with LS-based MMSE (\(+2.1\%\),  due to pilot-estimation overhead), and approximately \(311.7\)~s with the proposed ND, about \(1\%\) faster than LS-based MMSE while requiring no explicit CSI estimation.

\section{Conclusion}
We studied DeepSC over THz channels and showed that AWGN-trained DeepSC generalizes well with effective full-CSI MMSE compensation. A CSI-free system is proposed by replacing MMSE with a lightweight pilot-free ND while keeping AWGN training of DeepSC fixed, enabling fully data-driven semantic transmission without pilots or explicit CSI. The proposed ND outperforms MMSE with full, pilot-based, and noisy CSI, remains robust to \(\pm50\)~MHz offsets, and motivates future extensions to THz wideband MIMO and improved CSI-free semantic detectors.

\bibliographystyle{unsrt} 
\bibliography{references}    

\end{document}